\documentclass[aps,reprint,prd,showpacs,amsmath,amssymb,nofootinbib]{revtex4-1}
\usepackage{graphicx}
\usepackage{color}
\usepackage{times}
\usepackage{inputenc}
\usepackage{bm}
\usepackage{multirow}
\usepackage{float}
\usepackage{url}
\usepackage{natbib}
\usepackage{tensor}
\usepackage[colorlinks=true,citecolor=blue,urlcolor=blue,linkcolor=red]{hyperref}
\usepackage[normalem]{ulem}

\def\msun{\,M_{\odot}}
\def\fm3{\;\text{fm}^{-3}}
\def\mev{\;\text{MeV}}
\def\gev{\;\text{GeV}}

\begin{document}
\title{The stability of color-flavor-locked quark matter and massive CFL quark stars}
\author{Wen-Li Yuan$^{1,2}$}
\email{ wlyuan@pku.edu.cn}
\author{Bikai Gao$^{3}$}
\email{bikai@rcnp.osaka-u.ac.jp}
% \author{Renxin Xu$^{1,2}$}
\affiliation{
$^1$School of Physics and State Key Laboratory of Nuclear Physics and Technology, Peking University, Beijing 100871, China;\\
$^2$Department of Astronomy, School of Physics, Peking University, Beijing 100871, China;\\
$^3$Research Center for Nuclear Physics (RCNP), Osaka University, Osaka 567-0047, Japan
}

\date{\today}

\begin{abstract}
Owing to the emergence of the attractive interaction between quarks, color superconductivity is expected to occur, with the color-flavor-locked (CFL) phase favored at high densities. This work investigates the absolute stability of beta-equilibrated CFL quark matter in bulk within the modified Nambu-Jona-Lasinio model, under color and electric charge neutrality conditions relevant to compact stars. Motivated by the possible existence of an ultra-low-mass central compact object in the supernova remnant HESS J1731-347 and the “mass-gap” secondary component in the GW190814 event, we systematically explore how vector repulsion, attractive diquark pairing, and nonperturbative vacuum effects influence the stiffness of CFL quark matter and its stability. Our findings suggest the existence of a physically viable region of parameter space, in which the CFL phase is the true ground state of strongly interacting matter, thereby theoretically supporting the scenario of self-bound quark stars. This configuration is not only consistent with current astrophysical constraints from NICER and LIGO/Virgo observations, but also provides a possible explanation for both the $\sim 2.6\msun$ secondary component in GW190814 and the ultra-low-mass compact object with $M = 0.77^{+0.20}_{-0.17}\msun$ in HESS J1731-347.
\end{abstract}

\maketitle

\section{Introduction}

Understanding the properties of strongly interacting matter at supranuclear densities remains a major challenge in QCD and compact-star physics~\cite{2001PhRvL..86.5223P,2008RvMP...80.1455A,2018RPPh...81e6902B}. It is generally believed that dense matter around the nuclear saturation density, $\rho_0 \sim 0.16~\fm3$, is composed of hadronic degrees of freedom, while at supranuclear densities a transition to quark matter may occur through chiral restoration or deconfinement~\cite{2008RvMP...80.1455A}. Besides terrestrial heavy-ion collision experiments, compact stars serve as a unique astrophysical laboratory for probing such cold and dense supranuclear matter.

One of the central questions related to the dense-matter equation of state (EoS) concerns the true ground state of baryonic matter, commonly known as the Bodmer-Witten hypothesis~\cite{1971PhRvD...4.1601B,1984PhRvD..30..272W}. After decades of investigation, both strange quark stars~\cite{1971PhRvD...4.1601B,1984PhRvD..30..272W} and up-down quark stars~\cite{2018PhRvL.120v2001H} have been proposed as alternative physical models for neutron stars (NSs)~\cite{1989PhLB..229..112C,1993PhRvD..48.1409C,1998PhLB..438..123D,1999PhLB..457..261B,2000PhRvC..62b5801P,2010MNRAS.402.2715L,2018PhRvD..97h3015Z,2018PhRvD..98h3013L,2021EPJC...81..612B,2022PhRvD.105l3004Y,2024PhRvD.110j3012Z,2024FrASS..1109463Z,2020PhRvD.101d3003Z,2021PhRvD.103f3018Z,2024Ap&SS.369...29S}. 
Potential candidates for quark stars include compact objects with gravitational masses below $1\msun$, such as the X-ray bursters GRO J1744-28~\cite{1998Sci...280..407C}, 4U 1728-34~\cite{1999ApJ...527L..51L} and SAX J1808.4-3658~\cite{1999PhRvL..83.3776L}. The recent analysis of an ultra-low-mass compact object in the supernova remnant HESS J1731-347, with a remarkably low mass of $M = 0.77^{+0.20}_{-0.17}\msun$ and a radius of $R = 10.4^{+0.86}_{-0.78}$~km~\cite{2022NatAs...6.1444D}, has renewed interest in quark-star scenarios. Although NS or hybrid star interpretations have also been put forward~\cite{2023PhRvD.107b3012T,2023PhLB..84438062L,2023PhRvC.108b5806B,2024PhRvD.109d3036H,2024PhRvD.109f3017L,2024PhRvC.109f5807G,2026SCPMA..6932011G,2025Univ...11..345K}, the
quark-star scenario can accommodate such a small and compact object much more naturally due to the self-bound nature at the surface. Motivated by this possibility, a number of recent studies have explored both strange and non-strange quark stars composed of normal quark matter within effective frameworks, such as the NJL-type models and the MIT bag model, demonstrating compatibility with current observational constraints~\cite{2023PhRvD.107k4015R,2023A&A...672L..11H,2024ApJ...967..159D}.

Another key theoretical ingredient in dense quark matter is color superconductivity, which originates from the attractive interaction between quarks in the color-antitriplet channel predicted by quantum chromodynamics (QCD). At intermediate densities, the two-flavor color-superconducting (2SC) phase may be realized, while at asymptotically high densities the color-flavor-locked (CFL) phase is expected to be the ground state of three-flavor quark matter\cite{1998PhLB..422..247A,1999NuPhB.537..443A,2008RvMP...80.1455A}.
Diquark pairing modifies the quasiparticle spectrum of quark matter and can substantially alter the EoS. As a consequence, the global properties of compact stars may also be affected. 
In this context, Ref.~\cite{2024ApJ...966....3Y} investigated the absolute stability of 2SC quark matter within an NJL-type model that self-consistently generates dynamical quark masses, incorporating both electric and color charge neutrality, but found that 2SC quark stars may not exist. Before this, Ref.~\cite{2002JHEP...06..031A} demonstrated the absence of a 2SC phase inside the compact stars within the MIT bag model for three-flavor quark matter including unpaired strange quarks~\cite{2003A&A...403..173L,2017PhRvC..95b5808F,2023PhRvD.108f3010O}. Along this line, the CFL quark stars have been investigated within the framework of the MIT bag model in several studies, neglecting the contribution from electrons and considering massless $ud$ quarks together with a fixed strange-quark mass.
Among these studies, Ref.~\cite{2023PhRvD.108f3010O} focused on explaining the low-mass compact object in HESS J1731-347 and demonstrated that the viable parameter space can be further constrained by requiring a large maximum mass, $M_{\rm TOV} \geq 2.6\msun$~\cite{2020ApJ...896L..44A}, with remaining consistent with the conformal bound~\cite{2022PhRvL.129y2702F}.

In this work, we employ an NJL-type model incorporating both
quark-antiquark vector interactions and diquark pairing to study
color-superconducting quark matter, with particular emphasis on
the CFL phase. Within this framework, we systematically examine
how these microscopic interactions affect the stability of CFL
matter, the resulting EoS, and the global properties of the
corresponding CFL quark stars.
The analysis imposes electric and color charge neutrality conditions required for a stable bulk quark-matter system, while dynamically generating the constituent quark masses and density-dependent pairing gaps in a fully self-consistent manner. Despite extensive studies of three-flavor color superconductivity, within NJL-type models, focusing on the fundamental aspects of the size of the pairing gaps and the free energy of the system~\citep{2001PhRvL..86.3492R,2002PhRvD..66i4007S,2004PhRvD..69a4014M,2005PhRvD..72e4024S,2005PhRvD..72f5020B,2005PhRvD..72c4004R}, the question of whether CFL quark matter can be absolutely stable, which requires the energy per baryon of CFL phase at zero pressure satisfies $E/A ~(\rm CFL) < E/A~(^{56}\mathrm{Fe}) \simeq 930\mev$, remains insufficiently explored. Here, we first identify the region of parameter space, in which CFL quark matter becomes absolutely stable within the NJL-type model and is capable of supporting self-bound quark stars.

The paper is organized as follows. In Sec.~\ref{sec:form}, we present the NJL-type model with vector and diquark interactions and outline the formalism for describing the CFL phase. In Sec.~\ref{results}, we show the numerical results for the dynamical quark masses, pairing gaps, and the resulting EOS, as well as analyze the conditions for absolute stability and the properties of CFL quark matter. Finally, we summarize our findings and discuss their astrophysical implications in Sec.~\ref{summary}.

\section{Formalism}\label{sec:form}
In this section, we briefly write down the frequently employed three-flavor NJL model to describe the effective interactions between quarks.

\subsection{The three-flavor NJL-type model} \label{3f NJL model}
The Lagrangian of the three-flavor NJL model reads~\cite{1992RvMP...64..649K,2005PhR...407..205B,2018RPPh...81e6902B}
\begin{equation}
\begin{aligned}
\mathcal{L}_{\rm NJL}^{3f} & =\bar{q}\left(\gamma^\mu p_\mu-m_q+\mu_q \gamma^0\right) q+\mathcal{L}^{4}+\mathcal{L}^{6}\, \\
\end{aligned}
\end{equation}
% \begin{widetext}
% \begin{equation} 
% \begin{aligned}
% \mathcal{L} & =\bar{q}\left(\gamma^\mu p_\mu-\hat{m}_q+\mu_q \gamma^0\right) q+ G \sum_{j=0}^8\left[\left(\bar{q} \tau_j q\right)^2+\left(\bar{q} i \gamma_5 \tau_j q\right)^2\right]-g_{\mathrm{V}}\left(\bar{q} \gamma^\mu q\right)^2 \\
% &+H \sum_{A, A^{\prime}=2,5,7}\left[\left(\bar{q} i \gamma_5 \tau_A \lambda_{A^{\prime}} C \bar{q}^T\right)\left(q^T C i \gamma_5 \tau_A \lambda_{A^{\prime}} q\right)+\left(\bar{q} \tau_A \lambda_{A^{\prime}} C \bar{q}^T\right)\left(q^T C \tau_A \lambda_{A^{\prime}} q\right)\right] -K\left(\operatorname{det}\left[\bar{q}\left(1+\gamma_{5}\right) q\right]+\operatorname{det}\left[\bar{q}\left(1-\gamma_{5}\right) q\right]\right)\, \\
% \end{aligned}
% \end{equation}
% \end{widetext}
where the four-fermion and six-fermion interaction terms are written as:
$\mathcal{L}^{4} =\mathcal{L}_\sigma^{4}+\mathcal{L}_{V}^{4} +\mathcal{L}_d^{4}$ and $\mathcal{L}^{6} =\mathcal{L}_\sigma^{6}$. Here, $q$ is the quark field operator with color, flavor, and Dirac indices. $\hat{m}_q$ is the quark current mass matrix and $\mu_q$ is the flavor-dependent quark chemical potential. 

The phenomenological scalar and vector four-quark interaction terms are
\begin{equation}
\begin{aligned}
\mathcal{L}_\sigma^{4}  = G \sum_{j=0}^8\left[\left(\bar{q} \tau_j q\right)^2+\left(\bar{q} i \gamma_5 \tau_j q\right)^2\right]\ ,
\end{aligned}
\end{equation}
and
\begin{equation}
\begin{aligned}
\mathcal{L}_{V}^{4}=-g_{V}\left(\bar{q} \gamma^\mu q\right)^2 ,
\end{aligned}
\end{equation}
which are responsible for the spontaneous chiral symmetry breaking and for generating universal repulsion between quarks, respectively. Here, $\tau_j (j=0, \ldots, 8)$ are the generators of the flavor-U(3) symmetries.
The third four-quark interaction term, $\mathcal{L}_d^{4}$, characterizes pairing quark scattering in the s-wave, spin-singlet, flavor- and color-antitriplet channel, giving rise to BCS-type pairing among quarks. It is expressed as
\begin{equation}
\begin{aligned}
\mathcal{L}_d^{4}= & H \sum_{A, A^{\prime}=2,5,7}\left[\left(\bar{q} i \gamma_5 \tau_A \lambda_{A^{\prime}} C \bar{q}^T\right)\left(q^T C i \gamma_5 \tau_A \lambda_{A^{\prime}} q\right)\right. \\
& \left.\quad+\left(\bar{q} \tau_A \lambda_{A^{\prime}} C \bar{q}^T\right)\left(q^T C \tau_A \lambda_{A^{\prime}} q\right)\right] \ ,
\end{aligned}
\end{equation}
where $\tau_A$ and $\lambda_{A^{\prime}}\left(A, A^{\prime}=2,5,7\right)$ are the antisymmetric generators of $\mathrm{U}(3)$ flavor and $\mathrm{SU}(3)$ color, respectively. This diquark interaction induces an attractive correlation between quarks. In the weak-coupling limit, such attraction can be understood as arising from single-gluon exchange~\cite{2008RvMP...80.1455A}. However, at the densities relevant for NSs, the strong nonlinearity of QCD renders a first-principles calculation intractable, and the interaction must therefore be treated phenomenologically.

The six-fermion interaction term is written as
\begin{equation}
\mathcal{L}_{\sigma}^{6} = -K\left(\operatorname{det}\left[\bar{q}\left(1+\gamma_{5}\right) q\right]+\operatorname{det}\left[\bar{q}\left(1-\gamma_{5}\right) q\right]\right)\ ,
\end{equation}
which corresponds to the instanton-induced axial anomaly in QCD and explicitly breaks the $U(1)_A$ axial symmetry of the QCD Lagrangian.

Under the mean-field approximation, the inverse single-particle propagator in Nambu-Gorkov space is given by
\begin{equation}
\begin{aligned}
S^{-1}(p)=\left(\begin{array}{cc}
\gamma^\nu p_\nu-\hat{M}+\hat{\mu} \gamma^0 & \gamma_5 \Delta_i R_i \\
-\gamma_5 \Delta_i^* R_i & \gamma^\nu p_\nu-\hat{M}-\hat{\mu} \gamma^0
\end{array}\right)\ .
\end{aligned}
\end{equation}
The effective quark mass matrix $\hat{M}$ is diagonal in flavor space, with components
\begin{equation}
\begin{aligned}
M_i=m_i-4 G \sigma_i+K\left|\epsilon_{i j k}\right| \sigma_j \sigma_k \ ,
\end{aligned}
\end{equation}
where $\sigma_i \equiv \langle \bar{q}_i q_i \rangle$ denotes the chiral condensate.
The three diquark pairing amplitudes are,
\begin{equation}
\begin{aligned}
\Delta_j=-2 H \left\langle q^T C \gamma_5 R_j q\right\rangle \ ,
\end{aligned}
\end{equation}
with matrices
\begin{equation}
\begin{aligned}
\left(R_1, R_2, R_3\right) \equiv\left(\tau_7 \lambda_7, \tau_5 \lambda_5, \tau_2 \lambda_2\right) .
\end{aligned}
\end{equation}
These condensates correspond to $(ds, us, ud)$ pairing channels, respectively. In the 2SC phase, only the $ud$ channel is nonvanishing, whereas in the CFL phase all three channels are present. 
The effective chemical potential matrix is
\begin{equation}
\begin{aligned}
\hat{\mu}=\mu_q-2 g_V n_q+\mu_8 \lambda_8+\mu_{\mathrm{Q}} Q\ ,
\end{aligned}
\end{equation}
which is flavor and color dependent. Here, $n_q = \langle q^\dagger q \rangle$ denotes the quark number density. 
At each momentum, the inverse propagator forms a $72 \times 72$ matrix, whose quasiparticle dispersion relations $\epsilon_j$ are obtained numerically. Each eigenvalue is fourfold degenerate, reflecting spin (twofold) and Nambu-Gorkov (twofold) degeneracies.

The thermodynamic potential density of quark matter is
\begin{equation}
\begin{aligned}
\Omega_{q}=&-2 \sum_{j=1}^{18} \int^{\Lambda} \frac{\mathrm{d}^3 \mathbf{p}}{(2 \pi)^3}\left[T \ln \left(1+e^{-\left|\epsilon_j\right| / T}\right)+\frac{\Delta \epsilon_j}{2}\right]\\
&+\sum_{i=1}^3\left[2 G \sigma_i^2+H \left|d_i\right|^2\right] -4 K \sigma_1 \sigma_2 \sigma_3-g_{V} n_q^2\ ,
\end{aligned}
\end{equation}
which is the negative of the pressure $P$.
Here, $\Delta \epsilon_j=\epsilon_j-\epsilon_j^{\mathrm{free}}$, with $\epsilon_j^{\mathrm{free}}$ the eigenvalues in the noninteracting quark system, and $\Lambda$ is an ultraviolet cutoff. %\green{(Do you think it is better to add some discussion on the three momentum cut-off or the proper time?)}
The electron contribution to the thermodynamic potential is 
\begin{equation}
\begin{aligned}
\Omega_e=-2 T \sum_{\lambda= \pm} \int \frac{\mathrm{d}^3 \mathbf{p}}{(2 \pi)^3} \ln \left(1+e^{-\left(E_e+\lambda \mu_{\mathrm{Q}}\right) / T}\right),
\end{aligned}
\end{equation}
with $E_e=\sqrt{\mathbf{p}^2+m_e^2}$. Then the total thermodynamic potential of the system reads
\begin{equation}
\begin{aligned}
\Omega= \Omega_q + \Omega_e\ .
\end{aligned}
\end{equation}
%--------------||--------------||--------------||--------------||
%--------------||--------------||--------------||--------------||

\subsection{Electric and color charge-neutrality conditions}

The stellar matter should be imposed of both 
electric and color charge neutrality conditions. For the diquark pairing structures discussed above, all color densities except $n_3=\left\langle q^{\dagger} \lambda_3 q\right\rangle$ and $n_8=\left\langle q^{\dagger} \lambda_8 q\right\rangle$ automatically vanish. Thus, including the term  $\mathcal{L}_{3,8}=\mu_3 q^{\dagger} \lambda_3 q+\mu_8 q^{\dagger} \lambda_8 q$ 
is sufficient to ensure color neutrality. The electric charge chemical potential couples to the charge density in the Lagrangian through a term $\mathcal{L}_{\mathrm{Q}}=\mu_{\mathrm{Q}}\left[q^{\dagger} Q q- q_e^{\dagger} q_e\right]$
where $Q=\operatorname{diag}(2 / 3,-1 / 3,-1 / 3)$ is the quark charge operator in flavor space.

The thermodynamic state of the system is determined by minimizing the free energy $\Omega$ with respect to the six condensates $\left\{\sigma_i, d_i\right\}$ and the quark density $n_q$, under the constraints of electric and color charge neutrality, which are imposed through
\begin{equation}
\begin{aligned}
n_j=-\frac{\partial \Omega}{\partial \mu_j}=0\ ,
\end{aligned}
\end{equation}
with $j=Q, 3,8$. The chiral and diquark condensates are determined by the six gap equations
\begin{equation}
\begin{aligned}
0=-\frac{\partial \Omega}{\partial \sigma_i}=-\frac{\partial \Omega}{\partial d_i}\ ,
\end{aligned}
\end{equation}
in which all derivatives are evaluated at fixed quark chemical potential. Finally, the quark density is given by
\begin{equation}
\begin{aligned}
n_q=-\frac{\partial \Omega}{\partial \mu_q}\ .
\end{aligned}
\end{equation}
To construct the EoS, these 10 coupled equations are solved self-consistently following the procedure outlined in Ref.~\cite{2005PhLB..615..102A}. From the resulting solutions, the energy density $\varepsilon\left(\mu_q, T\right)$ and the pressure $P\left(\mu_q, T\right)$ can be obtained.

Before proceeding with the calculations, the model parameters should be fixed. In the NJL-type model, the parameter set ${\Lambda, m_u, m_d, m_s, G, K}$ can be determined by fitting to QCD vacuum phenomenology. Specifically, we adopt the parameter set by Hatsuda and Kunihiro (HK)~\cite{1994PhR...247..221H}, given by $\Lambda = 631.4\mev$, $m_{u,d} = 5.5\mev$, $m_s = 135.7\mev$, $G\Lambda^2 = 1.835$, and $K\Lambda^5 = 9.29$, which is chosen to reproduce the experimental data of the pion decay constant $f_{\pi}=93\mev$, the pion mass $m_{\pi}=138.0\mev$, the kaon mass $m_{K}=496\mev$, and the $\eta'$ meson mass $m_{\eta'}=958\mev$. The remaining coupling constants, ${g_{V}, H}$, are not well constrained. Several studies suggested that $g_{V}$ is similar in magnitude to $H$ to explain lattice results~\cite{2011PhRvD..84e6010K,2013PhLB..719..131B}. Typically, one might assume $H/G=3/4$ for equal contributions from the quark-antiquark interaction channels and the Fierz-transformed diquark interaction channels~\cite{2005PhRvD..72e4024S,2015PhRvD..92j5030C}. In this work, we treat them as free parameters as previous works~\cite{2023PhRvD.108d3008Y,2024ApJ...966....3Y,2023Symm...15..745M,2025PhRvD.112h3041G}, assuming they are of the same order of magnitude as the couplings in the vacuum-fitted parameter set.

%-----------------|-----------------|-----------------|
\section{Results and discussion}
\label{results}

\subsection{Dynamically generated masses and pairing gaps}

%-----------------|-----------------|-----------------|
\begin{figure}
\centering
\includegraphics[width=\columnwidth]{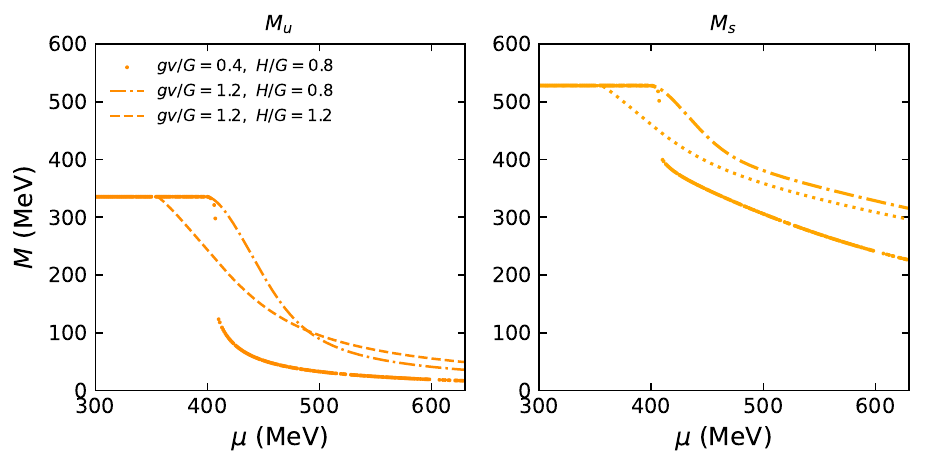}
% \vspace{0.1cm}
\includegraphics[width=\columnwidth]{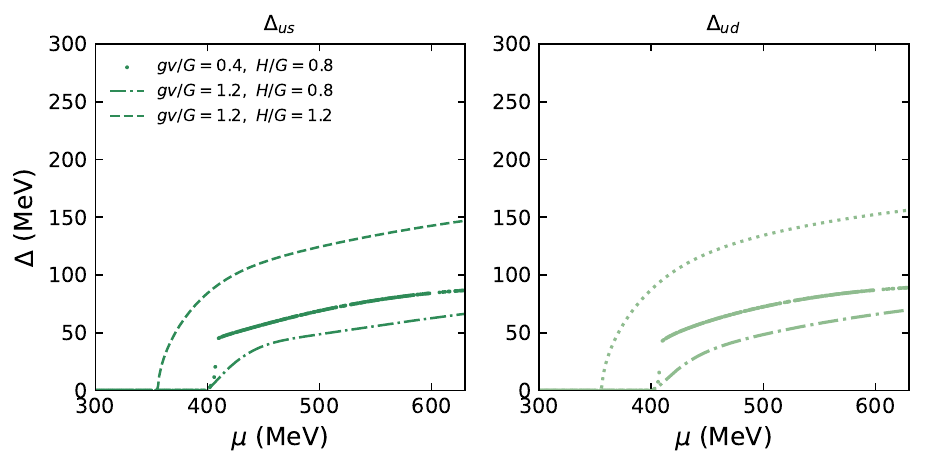}
\caption{The constituent quark mass $M_{u,s}$ (upper panel) and the diquark gap, $\Delta_{us} (\Delta_{ds})$ and $\Delta_{ud}$, (lower panel) as functions of the quark chemical potential for three parameter sets are shown: $(g_{V}/G=0.4$, $H/G=0.8)$, $(g_{V}/G=1.2$, $H/G=0.8)$, and $(g_{V}/G=1.2$, $H/G=1.2)$.
}
\label{fig:M Delta_miu}
\end{figure}
%-----------------|-----------------|-----------------|

%-----------------|-----------------|-----------------|
\begin{figure*}
\centering
\includegraphics[width=0.48\textwidth]{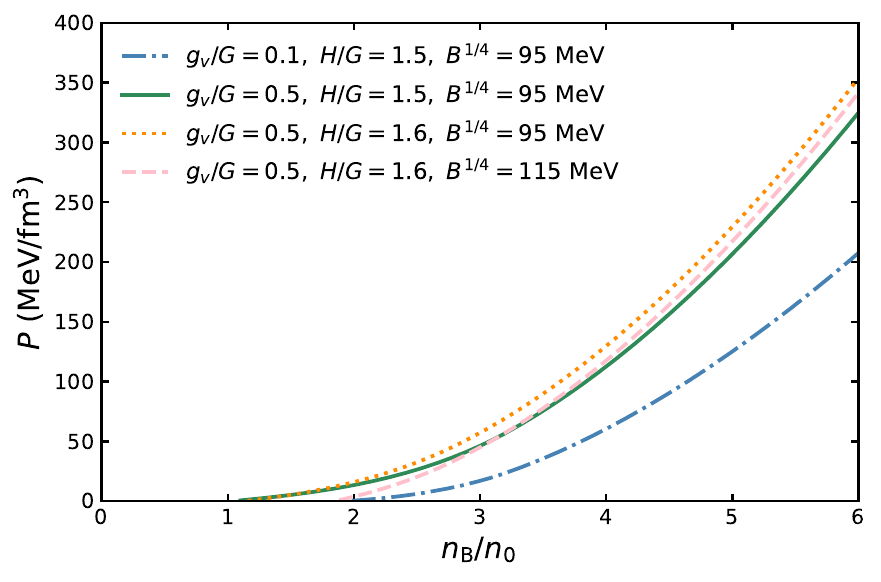}
\includegraphics[width=0.48\textwidth]{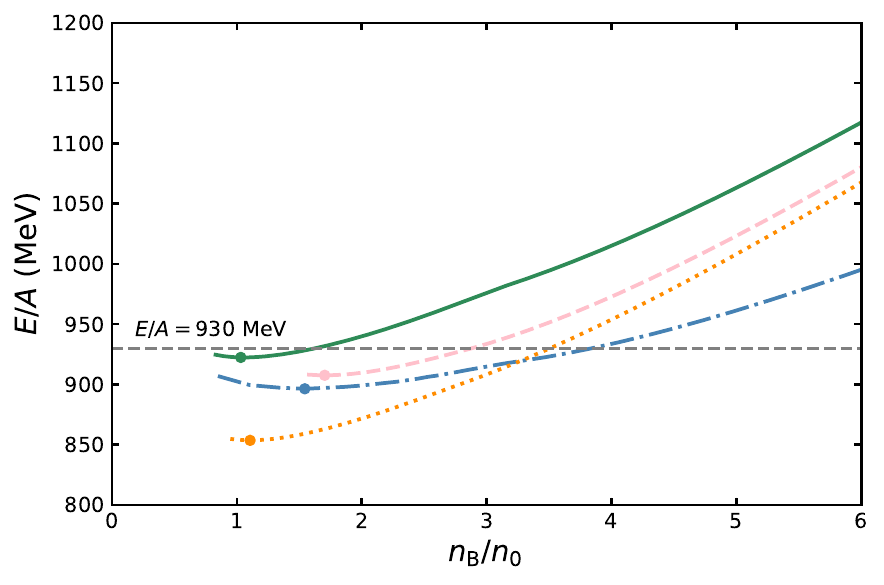}
\caption{Left panel: The pressure $P$ versus baryon number density $n_{\rm B}/n_{\rm 0}$ (in units of the nuclear saturation density $n_{\rm 0}$) for CFL quark matter within NJL-type model. Right panel: The energy per baryon as a function of $n_{\rm B}/n_{\rm 0}$ are shown, with the horizontal lines referring to the
energy per baryon of the most stable nuclei known, $E/A (^{56}\rm Fe)=930\mev$. The colored dots present the minimum values of $E/A$ for different parameter cases. The calculations are done for representative cases: $(g_{V}/G=0.1$, $H/G=1.5, B^{1/4}=95\mev)$, $(g_{V}/G=0.5$, $H/G=1.5,B^{1/4}=95\mev)$, $(g_{V}/G=0.5$, $H/G=1.6,B^{1/4}=95\mev)$ and $(g_{V}/G=0.5$, $H/G=1.6,B^{1/4}=115\mev)$.
}\label{fig:EOS and E/A}
\end{figure*}
%-----------------|-----------------|-----------------|

Figure~\ref{fig:M Delta_miu} displays the dynamical masses of up and strange quarks, $M_{u,s}$, in three-flavor quark matter, under the electric and color charge neutrality conditions for several representative parameter sets. The vacuum quark masses are $M_{u}^{\rm vac} = 324.6\mev$ and $M_{s}^{\rm vac} = 528.1\mev$, respectively. As the effective quark chemical potential exceeds the corresponding vacuum masses, chiral symmetry is progressively restored. The increase in the vector interacting strength $g_{V}/G$ modifies the effective chemical potential and smooths the density evolution, thereby suppressing abrupt rearrangements of the Fermi surface. As a result, when $g_{V}/G$ increases from $0.4$ to $1.2$, both the chiral and color-superconducting phase transitions evolve from first order to smooth crossovers. 
Diquark pairing arises from the attractive interaction between quarks near the Fermi surface. Increasing the strength of this interaction, from $H/G = 0.5$ to $1.2$, not only shifts the onset of the color-superconducting phase to lower chemical potentials but also enhances the pairing gap $\Delta$. In the lower panel of Fig.~\ref{fig:M Delta_miu}, for example, the diquark gaps reach $\Delta_{us} = 142.8\mev$ and $\Delta_{ud} = 152.5\mev$ at $\mu = 600.6\mev$ with $g_{V}/G = 1.2$ and $H/G = 1.2$, indicating that the magnitude of $\Delta$ is governed by both the strength of the attractive interaction and the density of states near the Fermi surface. At fixed quark chemical potential, the $ud$ pairing gap is slightly larger than the $us(ds)$ pairing gap due to the larger strange-quark mass, which induces a mismatch between the $u$ and $s$ Fermi surfaces, then suppresses the $us$ pairing.

%-----------------|-----------------|-----------------|
\subsection{The stability of stellar CFL quark matter}\label{sec_stability}

%-----------------|-----------------|-----------------|
\begin{figure}
\centering
\includegraphics[width=0.48\textwidth]{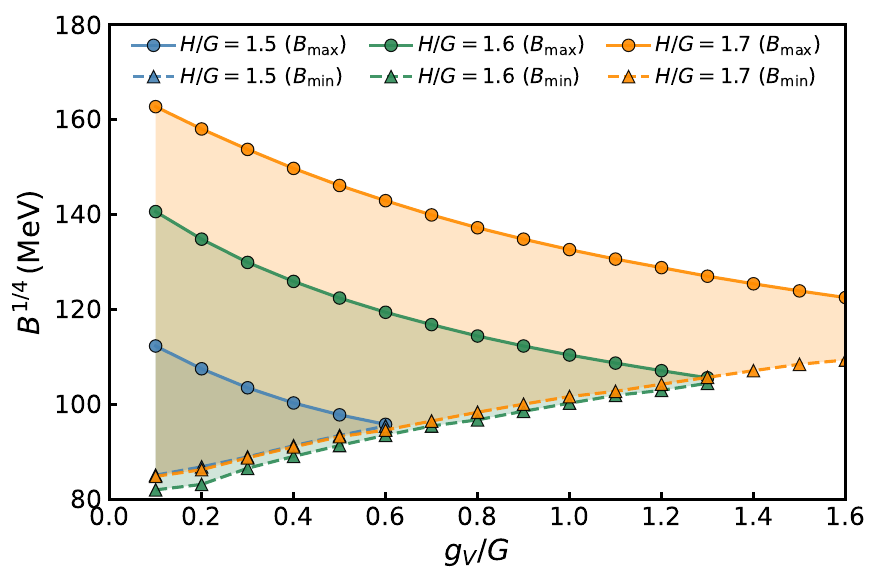}
\caption{Parameter space satisfying the minimum conditions for absolutely stable stellar CFL quark matter: $E/A < 930\mev$ and $n_{\rm B} > n_{0}$ at $P = 0$. The filled circles denote the maximum values of $B^{1/4}$ allowed by the condition $E/A < 930\mev$, while the triangles indicate the corresponding minimum values required by $n_{\rm B} > n_{0}$ at $P = 0$.
}\label{fig:space for B}
\end{figure}
%-----------------|-----------------|-----------------|

%-----------------|-----------------|-----------------|
\begin{figure*}
\centering
\includegraphics[width=0.48\textwidth]{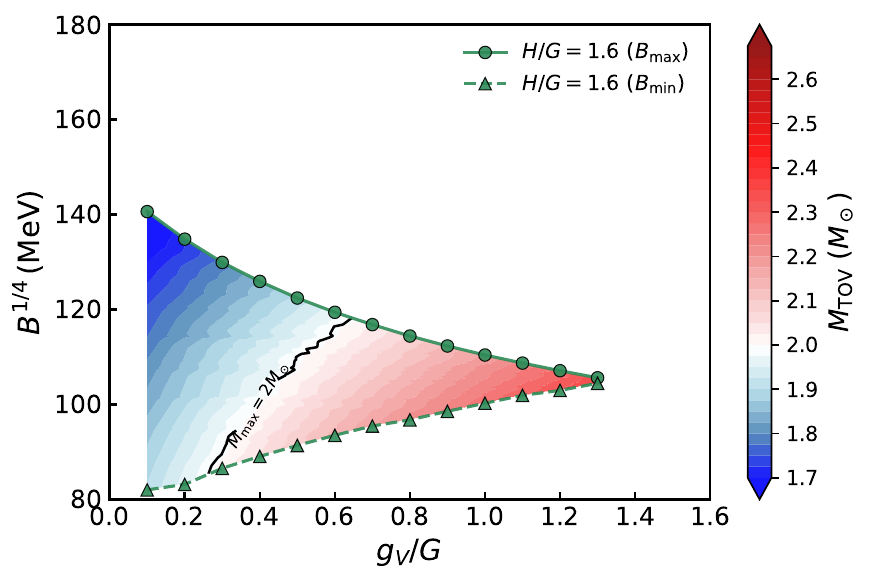}
\includegraphics[width=0.48\textwidth]{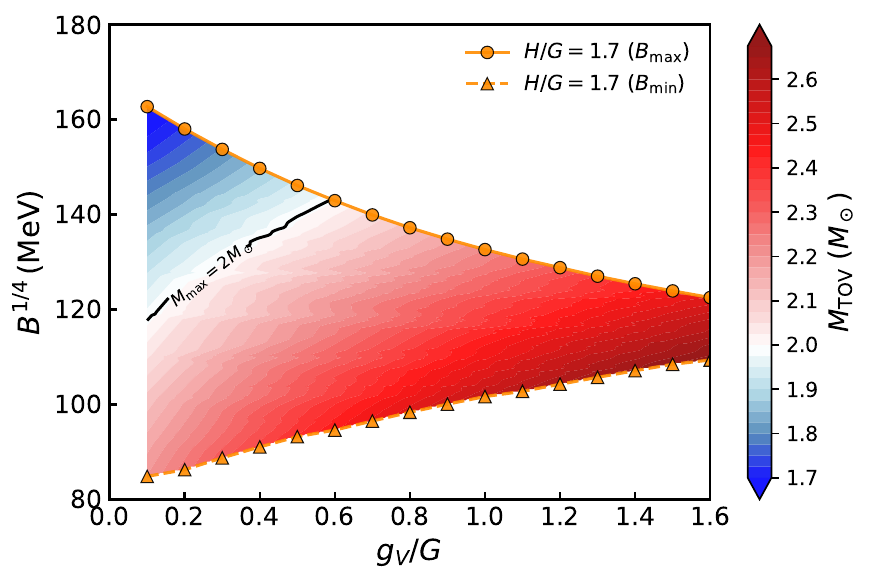}
\caption{Based on the absolutely stable parameter space shown in Fig.~\ref{fig:space for B}, this figure presents the predicted maximum mass of self-bound CFL quark stars as a function of the vacuum bag constant $B^{1/4}$ and the scaled vector coupling $g_{V}/G$, for $H/G = 1.6$ (left panel) and $H/G = 1.7$ (right panel). The white region separates configurations with maximum masses below $2\msun$ from those exceeding $2\msun$. The colored circles and triangles denote $B_{\max}$ and $B_{\min}$, respectively, with the same definitions as in Fig.~\ref{fig:space for B}.}\label{fig:mass scan}
\end{figure*}
%-----------------|-----------------|-----------------|

%-----------------|-----------------|-----------------|
\begin{figure*}
\centering
\includegraphics[width=0.97\textwidth]{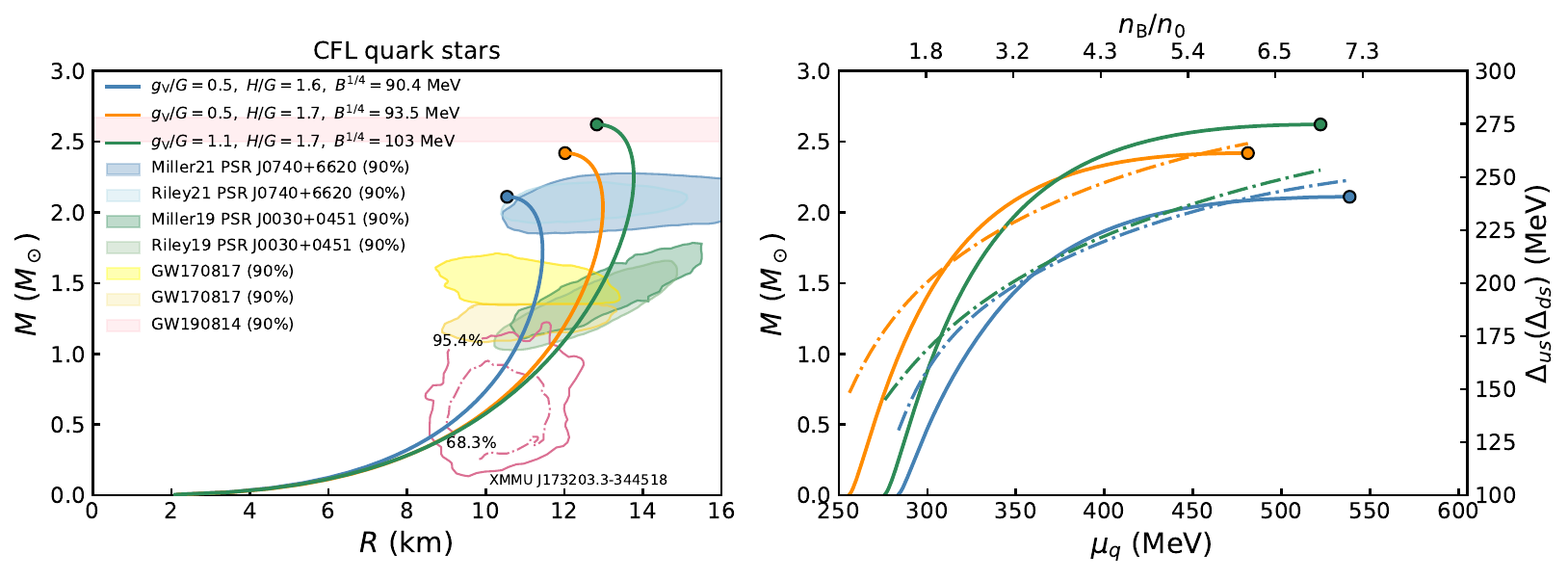}
\caption{Left panel: Mass–radius relations for several representative parameter sets within the NJL-type framework. The filled circles indicate the corresponding maximum masses for each sequence. Shown together are observational constraints from the NICER mission for PSR J0030+0451~\cite{2019ApJ...887L..24M,2019ApJ...887L..21R} and PSR J0740+6620~\cite{2021ApJ...918L..28M,2021ApJ...918L..27R}, and the binary tidal deformability constraint from LIGO/Virgo for GW170817~\cite{2017PhRvL.119p1101A,2018PhRvL.121p1101A}, all at the $90\%$ confidence level. The horizontal line indicates the mass measurement of GW190814's secondary component, $2.59_{-0.09}^{+0.08}\msun$ (at the $90 \%$ confidence level)~\cite{2020ApJ...896L..44A}, as well as a very
low-mass compact star, a
Central Compact Object named
XMMU J173203.3-344518
inside the supernova remnant
HESS J1731-347~\cite{2022NatAs...6.1444D}. Right panel: Stellar properties as functions of the quark chemical potential $\mu_q$ and $n_{\rm B}$. The solid curves show the gravitational mass, while the dashed curves represent the corresponding pairing gap $\Delta_{us}~(\Delta_{ds})$ for the same parameter sets as in the left panel. The lower horizontal axis denotes $\mu_q$, and the upper axis indicates the corresponding baryon number density $n_{\rm B}$.
}\label{fig:mass radius}
\end{figure*}
%-----------------|-----------------|-----------------|

To assess the stability of CFL quark matter and its ability to form self-bound quark stars, the energy per baryon at zero pressure must be lower than that of the most stable ion nucleus known, $E/A~(^{56}\mathrm{Fe}) \simeq 930\mev$~\cite{1971PhRvD...4.1601B,1984PhRvD..30..272W}. Within the NJL-type models, although the negative vacuum pressure $P(\mu=0 ; M)$ can be consistently calculated~\cite{2005PhR...407..205B}, the resulting description remains unsatisfactory due to the lack of confinement at zero density, as discussed in~\cite{2022PhRvD.105l3004Y}. In this work, we therefore treat $P(\mu=0 ; M)$ as a phenomenological parameter corresponding to $-B$ with $B$ serving the vacuum bag constant, which preserves the confinement of quarks.

Figure~\ref{fig:EOS and E/A} shows the pressure $P$ and the energy per baryon $E/A$ of CFL quark matter as functions of the baryon number density $n_{\rm B}/n_{\rm 0}$. The colored dots indicate the minimum values of the energy per baryon for different parameter sets, corresponding to zero pressure of the system. For the representative cases considered here, both the chiral and color-superconducting phase transitions are crossover. Increasing the vector interaction strength stiffens the EoS but renders the system energetically less favorable, as seen from the comparison between $g_{V}/G=0.1$ and $0.5$ at fixed $H/G=1.5$ and $B^{1/4}=95\mev$. 
In contrast, a stronger attractive interaction in the diquark channel enhances pairing near the Fermi surface, leading to larger gaps. The formation of a sizable diquark gap reorganizes the quasiparticle spectrum and suppresses low-energy excitations, thereby increasing the pressure response to compression and resulting in a stiffer EoS. At the same time, pairing lowers the free energy by forming correlated quark pairs, providing an additional binding contribution. Consequently, $E/A$ is reduced, rendering the system more energetically favorable and potentially absolutely stable. Such behavior is conducive to the formation of self-bound quark stars and to supporting large maximum masses consistent with observations. 
Increasing the vacuum bag constant $B$ softens the EoS of quark matter and shifts the onset of quark matter to higher baryon number densities at zero pressure. At the same time, it renders CFL quark matter energetically less favorable, as indicated by the increase in $E/A$. This behavior is clearly illustrated by comparing the cases $B^{1/4}=95\mev$ and $B^{1/4}=115\mev$ at fixed $g_{V}/G=0.5$ and $H/G=1.6$. 
To ensure that CFL quark matter satisfies the conditions for absolute stability and can form self-bound quark stars with finite surface density, the vacuum bag constant is subject to two basic constraints. First, $B^{1/4}$ must be sufficiently small such that the energy per baryon at zero pressure is below $930\mev$, ensuring absolute stability. Second, it cannot be too small; otherwise, the baryon number density at zero pressure (i.e., at the stellar surface) would not exceed the nuclear saturation density $n_{\rm 0}$, which is generally disfavored by the terrestrial experiments.

In addition to the stability condition for CFL quark matter, $(E/A)_{\rm CFL}<930\mev$, one must also ensure that ordinary nuclei remain stable against decay into deconfined two-flavor $ud$ quark matter, which requires $(E/A)_{\rm ud} > 930\mev$. In our previous work~\cite{2022PhRvD.105l3004Y}, we systematically examined the absolute stability of both $ud$ and $uds$ quark matter within a modified NJL model incorporating Fierz-transformed interactions, where the parameter $\alpha$ controls the relative weight of the Fierz-transformed channels. In the limit $\alpha=0$, this framework reduces to the standard two-flavor NJL model with vector interaction,
\begin{equation}
\begin{aligned}
\mathcal{L}_{\mathrm{NJL}}^{~2f} =&\,\bar{q}\left(i \gamma^{\mu}\partial_{\mu} - m + \mu \gamma^{0}\right)q\\
+&\,G\left[(\bar{q} q)^{2}+\left(\bar{q} i \gamma^{5}\tau q\right)^{2}\right]-g_{V}(\bar{q}\gamma^{\mu}q)^{2}\,. \label{eq: 2f NJL}
\end{aligned}
\end{equation}
Our previous results, shown in Fig.~5 of Ref.~\cite{2022PhRvD.105l3004Y}, demonstrated that bulk $ud$ quark matter described by Eq.~(\ref{eq: 2f NJL}) is not absolutely stable for any physically reasonable choice of parameters. A subsequent extension of this framework that further includes diquark pairing interactions~\cite{2024ApJ...966....3Y} likewise found no parameter space supporting absolutely stable 2SC quark matter. Taken together, these results guarantee that the constraint $(E/A)_{\rm ud,2SC} > 930\mev$ is automatically satisfied within the present model. Consequently, imposing only $(E/A)_{\rm CFL} < 930\mev$ is sufficient to delineate the physically allowed stability window for the CFL phase.

Based on the effects of repulsive vector interactions, attractive diquark interactions, and the vacuum pressure on the energy per baryon discussed above, we can determine the allowed range of $B^{1/4}$ for various values of $g_{V}/G$ and $H/G$ in CFL quark matter. The resulting stability window is shown in Fig.~\ref{fig:space for B}. For each fixed value of $H/G$, the filled circles trace the upper boundary $B^{1/4}_{\max}$ imposed by the absolute-stability requirement $E/A < 930\mev$, while the triangles trace the lower boundary $B^{1/4}_{\min}$ imposed by $n_{\mathrm{B}}(P=0) > n_{0}$. The region enclosed between these two curves defines the parameter space in which CFL matter constitutes the true ground state of strongly interacting matter and can therefore support self-bound quark stars. As $H/G$ increases, the allowed upper boundary becomes larger. This can be easily understood since a larger diquark coupling enhances the pairing gap and thereby lowers the minimum of $E/A$, so that absolute stability tolerates a
larger vacuum pressure.

\subsection{The CFL quark stars}

The EoS derived from the NJL-type model in Sec.~\ref{sec:form} uniquely determines the structure of the General Relativistic stellar configurations through integration of the Tolman-Oppenheimer-Volkoff (TOV) equations~\cite{1939PhRv...55..364T,1939PhRv...55..374O}.
In our calculations, only sufficiently strong attractive diquark interactions, $H/G > 1.5$, yield CFL quark stars with maximum masses exceeding $2\msun$. Accordingly, we present representative cases with $H/G = 1.6$ (left panel) and $H/G = 1.7$ (right panel) in Fig.~\ref{fig:mass scan}, where the parameter region capable of supporting $2\msun$ stars is highlighted in red. The color scale indicates the maximum mass $M_{\rm TOV}$ obtained for each combination of parameter sets $(g_V/G, B^{1/4})$ within the stability window discussed in Sec.~\ref{sec_stability}. 
The results demonstrate that stronger attractive interactions near the Fermi surface significantly enlarge the available parameter space capable of supporting massive stars with $M > 2\msun$, which can be understood from the enhancement of the diquark pairing gap. At fixed vacuum pressure and repulsive vector interaction, the diquark pairing not only lowers the free energy density but also increases the stiffness of the EoS, thereby enabling higher maximum masses. 

In Fig.~\ref{fig:mass radius}, the calculated mass-radius relations of CFL quark stars within the NJL-type framework are shown together with observational constraints from, such as, LIGO/Virgo~\cite{2017PhRvL.119p1101A,2018PhRvL.121p1101A} and NICER~\cite{2019ApJ...887L..24M,2019ApJ...887L..21R,2021ApJ...918L..28M,2021ApJ...918L..27R}.
As discussed above, decreasing the vacuum bag constant $B$, as well as increasing the vector coupling $g_{V}/G$ and the diquark coupling $H/G$, generally leads to a stiffer EoS, resulting in larger maximum masses for CFL quark stars. We find that a broad range of parameter combinations $(B^{1/4}, g_{V}/G, H/G)$ can simultaneously satisfy the constraints from LIGO/Virgo and NICER within the $90\%$ credible regions. In particular, such CFL quark star configurations can accommodate the $\sim 2.6\msun$ compact object identified as the secondary component in GW190814~\cite{2020ApJ...896L..44A}, serving as viable candidates for mass-gap compact objects. At the same time, the CFL quark star configurations can also consistently reproduce the inferred properties of the low-mass central compact object in the supernova remnant HESS J1731-347, $2.59_{-0.09}^{+0.08}\msun$ (at the $90 \%$ confidence level)~\cite{2020ApJ...896L..44A}. 
Note that previously Ref.~\cite{2023PhRvD.108f3010O} demonstrates, within the framework of the MIT bag model neglecting the electron contribution, that CFL quark stars can simultaneously explain the compact object in HESS J1731-347 and compact objects with masses comparable to or larger than the secondary component of GW190814 without violating the conformal limit.

The stellar mass increases monotonically with both the quark chemical potential $\mu_q$ and the baryon number density $n_{\rm B}$, as shown in the right panel of Fig.~\ref{fig:mass radius}. Correspondingly, the diquark pairing gap $\Delta$ reaches its maximum value at the stellar configuration with maximum mass, $M_{\rm TOV}$. For parameter set $g_V/G=0.5$, $H/G=1.6$, and $B^{1/4}=90.4~\mev$, the diquark pairing gap $\Delta_{us}$ is approximately $153.5\mev$ at the stellar surface and increases with baryon number density, reaching about $248.9\mev$ at the stellar center. Previous work in Ref.~\cite{2024PhRvL.132z2701K} employed model-agnostic constraints based on NS observations to estimate the allowed range of the color-superconducting gap $\Delta$ at densities beyond those directly accessible in NSs, where perturbative QCD calculations become reliable, and quark matter is expected to be in the CFL phase. They found that, at a baryon chemical potential $\mu_B = 2.6\gev$, the CFL pairing gap is constrained to be below $457\mev$ at the $95\%$ confidence level under conservative assumptions, and below $216\mev$ under more realistic assumptions. The latter constraint is broadly consistent with the range of pairing gaps obtained in the present NJL-type model calculations. We note here that, within the NJL-type framework considered here, the central quark chemical potential in all cases remains safely below the ultraviolet cutoff, $\Lambda = 631.4\mev$, thereby avoiding the internal inconsistency of the model associated with stellar central densities exceeding the density scale corresponding to the momentum cutoff mentioned in Ref.~\cite{2025PhRvD.111j3034G}. The corresponding central baryon number densities of CFL quark stars are found to lie in the range $n_{\rm B} \simeq 6.17-7.34~n_0$.

\section{Summary}\label{summary}

In this study, we investigated the possibility of absolutely stable CFL quark matter within the NJL-type framework including the vector and diquark interactions, employing a three-momentum cutoff regularization scheme to regulate the ultraviolet divergence. Our analysis highlights the significant impact of these interactions--particularly the attractive diquark interaction near the Fermi surface--on the stability of CFL quark matter. The existence of CFL quark star configurations is shown to be theoretically supported within the NJL-type model.

The increase of vacuum pressure and the repulsive vector interaction play roles opposite to that of the increasing attractive diquark interaction in determining the energy per baryon $E/A$ of CFL quark matter. In particular, a sufficiently strong diquark interaction near the Fermi surface is essential for lowering the free energy at zero pressure. Meanwhile, decreasing the vacuum bag constant $B$, as well as increasing the vector and diquark couplings $g_{V}/G$ and $H/G$, enhances the pressure response to compression, thereby stiffening the EoS. Within this NJL-type framework, we identify a region of parameter space that satisfies the absolute stability condition $E/A~(\rm {CFL}) < 930 \mev ~({}^{56}\mathrm{Fe})$, supporting the existence of self-bound CFL quark stars as an alternative to NSs. In particular, such CFL quark star configurations can simultaneously account for a wide range of astrophysical observations, from ultralow-mass $M = 0.77^{+0.20}_{-0.17}\msun$ compact objects~\cite{2022NatAs...6.1444D} to the $2.6\msun$ mass-gap compact object associated with GW190814~\cite{2020ApJ...896L..44A}, with remaining consistent with observational constraints from NICER and LIGO/Virgo. The difference in magnitude between the diquark gaps $\Delta{us}$ ($\Delta_{ds}$) and $\Delta_{ud}$ is approximately $10\mev$, arising from the mismatch between the Fermi surfaces of the light $u(d)$ quarks and the heavier $s$ quark. In the stellar core, the diquark gap can reach values as large as $\sim 280\mev$. In the future, within effective field-theory frameworks incorporating microscopic interactions, further investigations of the dynamical effects of the CFL phase, on compact-star cooling~\cite{2006NuPhA.777..497P}, r-mode instabilities~\cite{2000PhRvL..85...10M},  pulsar glitches~\cite{2022RPPh...85l6901A}, and other related high-energy astrophysical transient phenomena such as gamma-ray bursts~\cite{2005ApJ...632.1001O,2025PhRvD.111f3040S} will be important for revealing the physical properties of color-superconducting matter and the internal composition of compact stars.

\medskip
\acknowledgments
We thank Professor Ang Li for inspiring this idea. W.-L. Yuan is supported by the China Postdoctoral Science Foundation (Grant. No. 2025M773418). B.-K. Gao is supported by Grants-in-Aid for Scientific Research No. 26K17147.

\end{document}